\documentclass[journal]{IEEEtran}
\usepackage{optidef}
\usepackage{algorithmic,algorithm}

\usepackage{diagbox}
\usepackage{amsmath,amssymb}
\usepackage{enumerate,multirow,subfigure,graphicx,float,capt-of}
\usepackage[parfill]{parskip}
\usepackage[numbers,sort&compress]{natbib}
\usepackage{multirow}
\usepackage{caption}
\usepackage{bm}
\usepackage{tabulary}
\usepackage{lipsum,multicol}
\usepackage{multirow}
\usepackage{cases}
\usepackage[a4paper, total={6.5in, 8.5in}]{geometry}
\usepackage{threeparttable}
\usepackage{todonotes}
\usepackage{enumitem}
\usepackage{gensymb}
\usepackage{graphicx}
\usepackage{float}
\usepackage{amsthm}
\usepackage{cuted}
\usepackage{multicol}

\newtheorem{theorem}{Theorem}

\usepackage[switch]{lineno}

\begin{document}

\title{Submarine Cable Network Design for Regional Connectivity}
\author{Tianjiao~Wang, Zengfu~Wang,  Bill~Moran, Moshe~Zukerman, \IEEEmembership{Life Fellow,~IEEE}

\thanks{This work was supported in part by the City University of Hong Kong under Project 9667193, and in part by the Shenzhen Municipal Science and Technology Innovation Committee under Project JCYJ20180306171144091. \emph{(Corresponding author: Zengfu Wang.)}}
\thanks{Tianjiao Wang and Moshe Zukerman are with the Department of Electrical Engineering, City University of Hong Kong, Kowloon, Hong Kong (e-mail: tianjwang6-c@my.cityu.edu.hk; m.zu@cityu.edu.hk).}
\thanks{Zengfu Wang is with the Research \& Development Institute of Northwestern Polytechnical University in Shenzhen, Shenzhen 518057, China, and also with the School of Automation, Northwestern Polytechnical University, Xi'an 710072, China. (e-mail: wangzengfu@nwpu.edu.cn).}
\thanks{Bill Moran is with the Department of Electrical and Electronic Engineering, University of Melbourne, Melbourne, VIC 3010, Australia (e-mail: wmoran@unimelb.edu.au).}
} 

\maketitle

\begin{abstract}
This paper optimizes path planning  for a trunk-and-branch topology network in an irregular 2-dimensional manifold embedded in 3-dimensional Euclidean space with application to submarine cable network planning. We go beyond our earlier focus on the costs of cable construction (including labor, equipment and materials) together with additional cost to enhance cable resilience,  to  incorporate the  overall cost of branching units (again including material, construction and laying)  and the choice of submarine cable landing stations, where such a station can be anywhere on the coast in a  connected region.  These are important issues for the economics of cable laying and significantly change the model and the optimization process. 
We pose the problem as a variant of the Steiner tree problem, but one in which the Steiner nodes can vary in number, while incurring a penalty. We refer to it as  the \emph{weighted Steiner node problem}.  It  differs from the  Euclidean Steiner tree problem, where  Steiner points are forced to have  degree  three; this is no longer the case, in general, when nodes incur a cost.  We are able to prove that our algorithm is applicable to Steiner nodes with degree greater than three, enabling optimization of  network costs in this context. The optimal solution is achieved in polynomial-time using dynamic programming.
\end{abstract}


\title{SMT for coastline}

\section{Introduction}
\label{sec:Introduction}

\begin{figure}[htbp]
\centering
\includegraphics[width=7cm]{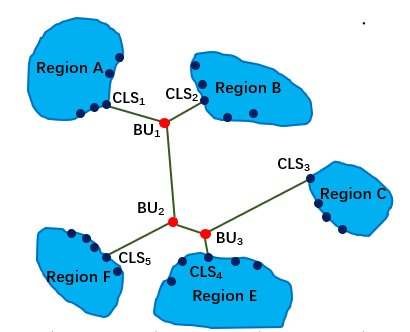}
\caption{Cable system interconnecting five regions.}
\label{fig:introduction}
\end{figure}

Submarine internet optical cables  play an important and crucial role in global communications, transmitting more than 99$\%$ of global Internet data~\cite{Submarinereport2020}. BY early 2021, there are more than 1.3 million kilometers submarine cables across the oceans and seas of the world~\cite{cablequestions} and,   according to~\cite{Submarinereport2020}, the capacity of submarine cables will increase $100\%$ by the end of 2023.
This paper studies the optimal  design of submarine cable systems. These systems typically have a trunk-and-branch tree topology as illustrated in Fig.~\ref{fig:introduction}. To be taken into account in their optimal design are  cable laying costs (including terrain slope, materials, alternative protection level, labor, and cable survivability),  installation costs of the submarine cable branching units (BUs), and the choice of  locations of  the submarine cable landing stations (CLSs). Our methods take  account of the 3-dimensional topography of the earth, by modeling it as an irregular 2-dimensional manifold in a 3-dimensional Euclidean space. Unlike earlier work, the  path planning optimization problem  considered here extends to the more realistic provision of region-to-region connectivity, as shown in Fig.~\ref{fig:introduction}, rather than just point-to-point. 

A submarine optical cable system is conveniently divided into two parts: underwater equipment and onshore equipment. The underwater equipment includes submarine optical cables, repeaters, and BUs, whereas the onshore equipment includes CLSs. Submarine BUs are used to fork cables, allowing traffic to be routed to (or merged from) two or more different locations thus permitting a diversity of connections for the cable system.  In most cases, a BU is  a Y-shaped cable connector, connecting three terminals~\cite{Basics}. However, in this paper, we do not exclude the possibility that a terminal which is connected to a BU can also be another BU. In this case, more than one BU will be used in a cable system,  and indeed, many cable systems use multiple BUs to allow multiple connections~\cite{Offshore}.
The cost of a BU is mainly determined by the number and complexity of fiber pairs in the cables~\cite{Basics}.  In 2002, the cost of a BU, was estimated to be between \$1 million and \$1.5 million~\cite{Paul}. A more recent estimate gives the cost as \$2 million~\cite{Dick}.

There are two main reasons to use BUs in a submarine cable network. 
\begin{enumerate}
\item  In practice, the total cable length should be as short as possible. By using BUs, the cable system can be constructed as a trunk-and-branch topology, providing a relatively small total length of cables. 
\item  In some areas, there is a requirement to transmit data to multiple destinations, but there are limitations to potential routes related to, for example, legal/licensing and seabed space restrictions~\cite{Elias}.
\end{enumerate}

When planning submarine cable routes, the choice of  location (or even existence)  of BUs needs to take account, not only of total cable length, but of  overall network cost. To this end,  account needs to be taken of, for example,  water pressure and depth, topography, etc., because under different seabed conditions, the type and cost of a BU varies. 

CLSs selection is another important factor in the construction of a submarine optical cable system  connecting several regions. Most CLSs are very close to the coast. In multi-point coastal systems, cost associated with CLSs  can become a significant component of total network cost. Large CLSs cost \$10-15 million while small CLSs are around \$5 million~\cite{Basics}. The location of CLSs can also affect the total length of  a cable network.
The specific locations of CLSs that serve regions/areas are usually carefully chosen,  taking into account the following issues~\cite{ye2018submarine,SUBMARINECABLENETWORK}.
\begin{enumerate}
\item   Marine zoning and compatibility with different  development and planning activities;
\item  Easy access to  landing points  by the submarine optical cable installation vessel for installation  and maintenance;
\item  Terrain risks for cables at landing points, for example, avoidance of steep rocky coasts and  areas of subsidence as well as steep, sandy or silty sea-floors;
\item  Environmental risks for cables such as storm surges and other marine disasters;
\item  Avoidance of  areas where  corrosion (chemical pollution) of optical cables is likely;
\item Avoidance of  areas of risk of cable damage by anchors and trawling nets.
\end{enumerate}
Such ideal locations are rare, and on the coastline of a given region, usually only a few locations are suitable. A CLS will sometimes be the shared landfall point for several cable systems~\cite{SUBMARINECABLENETWORK}.

The key novelty of this paper is the extension of the optimization problem of the submarine cable system from minimizing the total cable construction cost in a cable system  to minimizing the construction costs of the entire cable system, including, in addition to the cable construction costs,  the installation costs of BUs and CLSs. We pose the optimization problem as a variant of the Steiner minimum tree problem, on an irregular $2 \mathrm{D}$ manifold in a $3 \mathrm{D}$ Euclidean space and propose  polynomial-time dynamic programming for its solution. We call this \emph{the weighted  Steiner node problem}.

The contributions of this paper are as follows.
\begin{enumerate}
\item  In cable system construction optimization, the installation costs of BUs is included  in addition to minimization of  the total cable length. By optimizing cable network structure, we can determine the number of BUs that should be used in the cable network taking account of  the unit installation cost of a BU, which in turn depends on the environmental conditions in its location. 
\item In addition to optimizing the network based on the limited assumption that all BUs are Y-shaped (have three branches), our algorithm also solves the cable network optimization problem with multiple (more than three) branch BUs. We introduce a theorem in Section~\ref{sec:method} that proves that our algorithm is applicable to Steiner nodes with degree of three or more.
\item   Cable connections between regions, rather than between locations become the focus of the optimization,  which implies that our destinations are multiple sets of points. Such a cable network connects  regions rather than specified coastal locations.
We consider a range of CLS construction alternatives within a region. Our algorithm can select the best CLS in each region according to their locations and construction cost, so that the total construction cost of the cable network is minimized.
\end{enumerate}

The remainder of this paper is organized as follows. In Section~\ref{sec:rel_res}, we review related  path planning research in cable networks, including point-to-point path planning and cable network path planning algorithms. Section~\ref{sec:prob_form} focuses on the  modeling of our cable network optimization problem. In Section~\ref{sec:method}, we propose a method to solve the problem. The performance of our method is demonstrated by several numerical simulations based on realistic examples in Section~\ref{sec:num_res}. Finally, we conclude the paper in Section~\ref{sec:conclusion}. 

\section{Related Research}
\label{sec:rel_res}
In the current extensive literature on submarine cable path planning, we concentrate on optimization of the following three relevant network topologies in the context of submarine cable path planning: (1)  point-to-point~\cite{wang2021jlt1, wang2019path,wang2017multiobjective,zhao2016route}; (2) point-to-area/existing cables~\cite{wang2019cost}; and (3) cable networks connecting multiple~(more than two) points~\cite{wang2020optimal, wang2021optimal}.

In practice, the industry generally uses a manual approach, based on expert experience, for the path planning of submarine cables~\cite{eramo2018processing}. Based on data available on the relevant region, planners determine the cable path by checking carefully all the relevant details along the path and comparing between  various alternatives. This approach cannot guarantee an optimal path and can be expensive in human resources. 

Currently, MakaiPlan, an industry-leading submarine cable route planning software is available on the market. For a pair of nodes to be connected, MakaiPlan chooses the shortest distance along a great circle route in a series of rhumb lines, providing  a rough estimate of the geodesic. MakaiPlan only provides path planning of submarine cables connecting two nodes, and provides no means to optimize, taking account of  risk factors,  submarine cable networks with trunk-and-branch and mesh architectures: commonly used topologies in submarine cable systems. 

In~\cite{zhao2016route}, using the Dijkstra's algorithm, the authors proposed a raster-based method for path planning to obtain the cable path with least cost, specifically, minimizing the weighted sum of the seismic risk and the cable length (cost). A disadvantage of this method is that a path must traverse either a diagonal or a lateral edge between neighboring cells. Additionally, they only considered point-to-point cable path planning.

The fast marching method (FMM) is a path planning approach that avoids the weakness of the Dijkstra’s algorithm by solving the Eikonal equation as the grid size of the triangulated manifold approaches zero~\cite{sethian1999level}. In~\cite{wang2017multiobjective}, Wang \emph{et al.} presented an FMM-based approach to solve the path planning problem on the earth’s surface.
More recently, in~\cite{wang2020optimal,wang2021optimal,wang2019cost}, FMM was  applied to cable path planning, designing a cable path from a start node to an existing cable network. Their objective was to lay a cable from node $A$ to a certain end location $B$ on an existing cable at a minimal cost. They considered three scenarios for the destination:  an arbitrary location on an existing cable with the addition of a new BU, an installed BU or a landing station; a landing station. The proposed method can be applied to cable path planning from one node to a set of nodes. Wang {\it et al.}~\cite{wang2020optimal} proposed an FMM-based method to achieve a trunk-and-branch tree network topology for submarine cable systems that connect multiple landing stations. That work considered the optimization of the  trunk-and-branch tree network problem as a Steiner minimum tree~(SMT) problem, and based on dynamic programming, obtained the minimum tree on an irregular $2 \mathrm{D}$ manifold in a $3 \mathrm{D}$ Euclidean space for a given topology in polynomial time. However, it did not take account of installation costs of BUs and, in that paper, the CLS location in every region was uniquely specified.
In~\cite{wang2021optimal}, we proposed a method based on FMM and Integer Linear Programming to optimize a tree-topology network design for a submarine cable system that connects multiple stations. We also proposed a lightweight heuristic algorithm to solve this problem. We considered the optimization of the cable network problem as a minimum spanning tree~(MST) problem. The objective was to construct a tree-topology cable network without additional Steiner nodes at minimal cost. In that work,  the installation costs of BUs, and location choices of CLSs were not considered.

The SMT problem in $d$-dimensional Euclidean space is a well-known NP-hard combinatorial optimization problem. Many authors have contributed ample research on this problem and proposed many solutions~\cite{gilbert1968steiner, warme1998spanning,winter1997euclidean,caleffi2014solution,caleffi2014solution,aharoni1998restricted}.
 Gilbert and Pollak~\cite{gilbert1968steiner} proposed a method to solve the SMT problem in $d$-dimensional Euclidean space. In their algorithm, the first step is to enumerate all Steiner topologies and then calculate the relative minimal tree (RMT) corresponding to each topology. Their algorithm is very computationally intensive, as the number of Steiner topologies grows super-exponentially with the number of terminals.  As a result, it can only be applied to very small instances of the SMT problem. The GeoSteiner algorithm~\cite{GeoSteiner} can solve a typical SMT problem with a large number of terminals in the 2-dimensional Euclidean plane by using geometrical properties (such as node degree and the angle conditions). However, GeoSteiner heavily utilizes the properties of the Euclidean plane, and these geometrical properties do not translate to higher dimensions and irregular 2-dimensional manifolds in $\mathbb{R}^{3}$.  Smith~\cite{smith1992find} proposed an implicit enumeration scheme to generate full Steiner topologies by using branch and bound~($\mathrm{B\&B}$). Smith's $\mathrm{B}\& \mathrm{B}$ algorithm consists of a tree generation step and a numerical optimization step. In the $\mathrm{B}\& \mathrm{B}$ algorithm, there is a continuously updated maximum value. Spanning trees that exceed the maximum value are eliminated, thus reducing the computational complexity. The metric in~\cite{smith1992find} is  Euclidean (albeit $d$ dimensional), so it is not applied directly to our case of an irregular  2-dimensional manifold $\mathbb{M}$ in $\mathbb{R}^{3}$.  Fampa \emph{et al.}~\cite{fampa2008improved} improved Smith's $\mathrm{B\&B}$ algorithm by using a conic formulation for the subproblem of locating the Steiner nodes, but the general Euclidean SMT problem in cases other than the 2-dimensional plane remains challenging. The work  discussed above on the SMT problem in general Euclidean space cannot be directly used in our problem of finding a minimal cost tree in a manifold. 
Approximate solutions to the SMT problems have been applied to communication
networks design in~\cite{caleffi2014solution,sun2020physarum}. 
Caleff {\it et al.} ~\cite{caleffi2014solution} used an algorithm based on a BioNetwork of a unicellular
organisms for the SMT problem, and 
Sun {\it et al.}~\cite{sun2020physarum} proposed an algorithm based on Physarum-inspired algorithm for the SMT problem. 

Unlike the available approximate solutions, we seek an optimal solution for the problem under consideration. 
The closest study to our work is the above-mentioned work~\cite{wang2020optimal}, which provided a polynomial time complexity numerical algorithm, based on dynamic programming, to find the RMT on the manifold. But the work in~\cite{wang2020optimal}  did not consider the cost of Steiner nodes and terminal nodes in the network which is what is achieved in the present paper.

\section{Problem Description and Modeling}
\label{sec:prob_form}

In this section, we pose, mathematically the  optimal cable network design problem,  called the \emph{weighted Steiner node  problem}. To address this problem, a 2-dimensional manifold in 3-dimensional space is used to model the surface of the earth on which we aim to minimize the total cost of a submarine cable system. 

Let $\mathbb {D}$ denote a bounded closed region on the earth’s surface modeled as a 2-dimensional manifold, where any pair of nodes in this region can be connected by a path on the 2-dimensional manifold. As in~\cite{wang2017multiobjective, wang2018application, wang2019terrain}, to  represent the earth's surface with a reasonable level of fidelity, we use a triangulated piecewise-linear 2-dimensional manifold $\mathbb{M}$ to model the region $\mathbb{D}$. The ``smoothly" continuous area $\mathbb{D}$ is  rendered as a triangulated manifold $\mathbb{M}$, each point $\mathbf{x}$ on $\mathbb{M}$ being represented by 3-dimensional coordinates $(x, y, z),$ where $z=\xi(x, y)$ is the altitude of $(x, y)$. For a cable represented by a Lipschitz continuous~\cite{eriksson2013applied} curve $\gamma$ connecting two nodes in $\mathbb{M}$, the total construction cost (based on the length) of the cable $\gamma$ is denoted by $\mathbb{H}(\gamma)$, 
as shown in Eq.~(\ref{equ:hfunction}),
where  $f(\mathbf{x})$ is the laying (construction) cost per unit length of the cable passing through point $\mathbf{x}$ on the cable.

\begin{equation}
\label{equ:hfunction}
 \mathbb{H}(\gamma)=\int_{0}^{l(\gamma)} f(\gamma(s)) d s.   
\end{equation}
The value of $f(\mathbf{x})$ is discussed in~\cite{wang2017multiobjective}, where  an additive assumption of cable laying cost is made, accommodating  several factors, including soil type, elevation, labor, licenses, protection level and the direction of the path~\cite{wang2019cost}.

Let $\mathbf{R}_1, \mathbf{R}_2, \ldots, \mathbf{R}_{R} \subseteq \mathbb{D}$ ($R$ is the number of regions) be the destination regions to be connected via a cable network with a trunk and branch topology. Note that, although in this paper each $\mathbf{R}_i$ represents a region, our solutions are applicable also to the case where some or all of the $\mathbf{R}_is$  correspond to a specific point.  

The network may include new nodes called \emph{Steiner nodes} (representing BUs in the cable network) to reduce total cable length of the network and influence the overall  construction  cost. As discussed in Introduction, the construction costs of the cable network, in our situation,  includes the costs of submarine cables, BUs and CLSs, so decisions  on the number of BUs in the network and  locations of CLSs (if there is a choice) are needed.


We denote by $\mathbf{b}_i$ the BU installation cost  at  the grid node $\mathbf{x}_i$ in $\mathbb{M}$. When a Steiner node has the same location as a terminal node,  $\mathbf{b}_i=0$. This means that, since a Steiner node represents a BU, if a Steiner node is included in a terminal node, the elimination of the Steiner node implies a potential savings as  the BU is not required, and the number of BUs is reduced by one. 

If BUs in the cable systems have no branch number constraints, there is another case in which the cost of Steiner nodes are ignored: Two or more 3-branch BUs have the same location. Then they can be taken to represent a single BU which the number of branches is the addition of the number of branches of each combined BU and subtract $2$, so we  count the BU cost (may depend on the number of branches) once.

 Because of the widespread usage  of  3-branch BUs in the industry, we demonstrate in realistic cable networking scenarios that our algorithm can restrict the solutions to just 3-branch BUs. Additionally,   one realistic example of a cable system with $K$-branch ($K \geq 3$, $K \in \mathbb{N}^+$) BUs is demonstrated in Section~\ref{subsec:BUbranchmul}. 

\begin{figure}[htbp]
    \centering
    \includegraphics[width = 0.9\columnwidth]{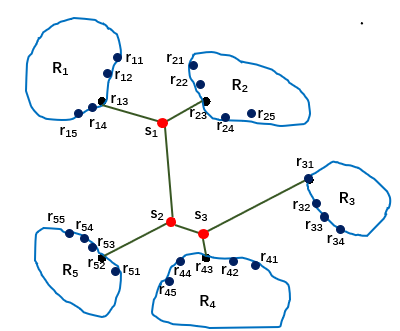}
    \caption{SMT for regions connecting.}
    \label{fig:region}
\end{figure}

Another consideration in cable network optimization is the choice of CLS locations. Fig.~\ref{fig:region} gives a simple and clear explanation of this problem. For this problem, more than one potential site can be the location of  a CLS in each region $\mathbf{R}_i$. Let $\mathbf{r}_{i1}, \mathbf{r}_{i2}, \ldots, \mathbf{r}_{iL_i} \in \mathbf{R}_i$ be the potential locations for CLSs in region $\mathbf{R}_i$.  As discussed in Introduction, the construction costs of the CLSs depends on many factors. In each region, several potential locations are available for CLS of the region. Different locations have different construction cost and will also result in different lengths of cables in the network. 

\section{Methodology and Algorithm}
\label{sec:method}
Our cable network optimization problem on the triangulated manifold $\mathbb{M}$ is clearly NP-hard since it generalizes the SMT problem in the 2-dimensional Euclidean plane. 

We think of tree  topologically $\mathcal{T}$ as the combined structure of points and edges in terms of  topological rather than metric entities, but including the labelling of some nodes as \emph{Steiner nodes} and the others as \emph{terminal nodes}, as well as  specifying  which terminal/Steiner node pairs are connected by edges. Such a topology $\mathcal{T}$ is called a \emph{Steiner topology} if the degree of each Steiner node  is equal to three and the degree of each terminal node is three or less~\cite{fampa2008improved}.
We use the generic term \emph{topology} of a tree to mean that there are no restrictions on the degrees of Steiner nodes and terminal nodes, though again with a labelling of nodes as Steiner nodes or terminal nodes. General topologies may include, in particular,  some Steiner nodes with more than three branches. For given $N$ terminal nodes, Steiner topologies form a subset of the set of  all tree topologies.

\begin{figure}[htbp]
    \centering
    \includegraphics[width = 8cm]{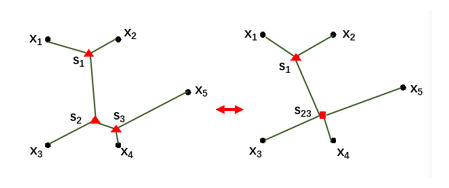}
    \caption{A 4-branch BU derived from two 3-branch BUs.}
    \label{fig:multibranches1}
\end{figure}

\begin{figure}[htbp]
    \centering
    \includegraphics[width = 8cm]{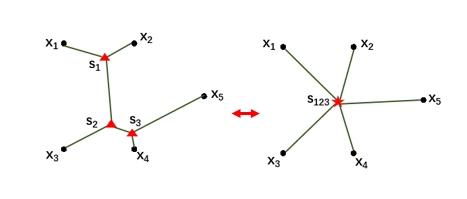}
    \caption{A 5-branch BU derived from three 3-branch BUs.}
    \label{fig:multibranches2}
\end{figure}

We extend these ideas to cover topologies in which there are  $K$-degree ($K \geq 4$, $K \in \mathbb{N}^+$) Steiner nodes by regarding such nodes as combinations of  linked $3$-branch Steiner nodes, with zero-cost (and, thereby,  zero length) edges joining them. In this way, a combination of two 3-degree Steiner nodes will produce a 4-degree Steiner node,  three  linked 3-degree Steiner nodes  will produce a 5-degree Steiner node, and so on. As Fig.~\ref{fig:multibranches1} shows, the new Steiner node $\mathbf{s}_{23}$ is a combination of the Steiner nodes $\mathbf{s}_{2}$ and $\mathbf{s}_{3}$; in Fig.~\ref{fig:multibranches2}, the new Steiner node $\mathbf{s}_{123}$ is a combination of the Steiner nodes $\mathbf{s}_{1}$, $\mathbf{s}_{2}$ and $\mathbf{s}_{3}$. 

If a tree involves a mix of $K$-degree ($K \geq 4$, $K \in \mathbb{N}^+$) Steiner nodes and 3-degree Steiner nodes (there may be no 3-degree Steiner nodes so it could consist of only $K$-degree Steiner nodes), we can take any $K$-degree Steiner node and split it into $(K-2)$ 3-degree Steiner nodes. Note that the order in which we do this is irrelevant. 
For terminal nodes with more than three branches, we can add more virtual BUs and zero-cost edges~\cite{smith1992find}. The topology of this tree then becomes a Steiner topology. The reverse  conversion is also valid. Therefore, if we know all Steiner topologies then we know all topologies. 

From the above discussion, we are able to articulate the following theorem.
\begin{theorem}
For a given graph with $N$ nodes, a general topology with $K$-degree ($K \geq 3$, $K \in \mathbb{N}^+$) Steiner nodes can be derived by the process described above from a Steiner topology  (with only $3$-degree).
\end{theorem}

Our new dynamic algorithm for the minimal cost network problem is based on solving the Steiner tree problem with  Steiner nodes having three branches, that is, the SMT problem. But our algorithm can also be applied to the case where the tree topology includes $K$-degree ($K \geq 3$, $K \in \mathbb{N}^+$) Steiner nodes. 

A Steiner topology with $N$ terminal nodes is a \emph{full} Steiner topology if there are $N-2$ Steiner nodes, and the degree of each terminal node is equal to one. The number of Steiner nodes in a Steiner topology can only be reduced by coalescing  Steiner nodes with terminal nodes,  while for generic topologies, the number of Steiner nodes can also be reduced by coalescing Steiner nodes with other Steiner nodes to create higher branched Steiner nodes.

Given $N$ terminals $\mathbf{x}_{1}, \mathbf{x}_{2}, \ldots, \mathbf{x}_{N} \in \mathbb{M}$, let $\mathbf{N} = \{1,2,\ldots, N\}$ be the index set of terminals, $\mathbf{S}= \{N+1, N+2,...,N+M \}$ be the index set of Steiner nodes~($M\leq N-2$), and  $\mathbf{V}=\mathbf{N} \cup \mathbf{S}$.
Given a topology $\mathcal{T}$ derived by Smith's $\mathrm{B} \& \mathrm{B}$ algorithm, let $\mathbf{E}=\mathbf{E}_{1} \cup \mathbf{E}_{2}$ be the set of all edges, i.e., $\mathcal{T}=(\mathbf{V}, \mathbf{E})$, where $\mathbf{E}_{1}=\{(i, j) | i \in \mathbf{N}, j \in \mathbf{S}\}$ and $\mathbf{E}_{2}=\{(i, j) | i \in \mathbf{S}, j \in \mathbf{S}\}$. 
We aim to find the coordinates of Steiner nodes: $\mathbf{x}_{N+1}, \mathbf{x}_{N+2}, \ldots, \mathbf{x}_{N+M} \in \mathbb{M}$ and the paths $\Gamma=\{\gamma(e) | e \in \mathbf{E}\}$ (i.e., the geodesics corresponding to the edges in $\mathbf{E}$). 

We impose the constraint that all terminals and all potential Steiner nodes are  grid nodes $\mathbf{x}_i$ of $\mathbb{M}$, $\mathbf{x}_i \in \mathbb{M}$ (we are only given the coordinates of grid nodes of $\mathbb{M}$ in practice). By applying FMM, we  are able to find the shortest path between every terminal node and  grid node in $\mathbb{M}$.

For a cable segment $\gamma_{i j}$ similar to~\cite{wang2020optimal} that connects two grid nodes $\mathbf{x}_{i}, \mathbf{x}_{j} \in \mathbb{M}$, let $c\left(\mathbf{x}_{i}, \mathbf{x}_{j}\right)=\int_{\mathbf{x}_{i}}^{\mathbf{x}_{j}} f(\mathbf{x}(s)) d s$ denote the cumulative weighted cost (here after called ``cost" for brevity) over the cable path $\gamma_{i j}$. Note that the cable is not required to traverse edges of triangles of $\mathbb{M}$; we assume that the cable can pass through the interior of triangles of $\mathbb{M}$ and FMM optimizes the path accordingly.  

\subsection{Minimal Cost Steiner Trees with  Steiner Nodal Cost}
\label{subsec:SMT_BU}

The mathematical description $\mathcal T$  of a physical cable network comprises the locations of all Steiner nodes, all terminal nodes, and all paths of cables.
 We denote  the total cost of the physical cable network $\mathcal T$  by $\Psi(\mathcal T)$.  At this stage, we do not include the cost of CLSs. Eq.~(\ref{equ:mincostofnetwork}) shows the problem of optimizing the cost of a Steiner tree ignoring  the cost of Steiner nodes. 
\begin{equation}
\begin{aligned}
\label{equ:mincostofnetwork}
\min\limits _{X}\!\Psi(\mathbf{X}, \Gamma) \! = \! \min\!\!\bigg (\sum\limits_{j \in \mathbf{S}} \bar{c}\left(\mathbf{x}_{j}\right)\!+ \!\!\!\!\!\!\!\! \sum\limits_{(i, j) \in \mathbf{E}_{2}} \!\!\!\!\!\min\limits _{\gamma_{i j}} c\left(\mathbf{x}_{i}, \mathbf{x}_{j}\right)\!\!\!\bigg)    
\end{aligned}
\end{equation}
where $\bar{c}\left(\mathbf{x}_{j}\right)= \sum\limits_{{i \in \mathbf{N},(i, j) \in \mathbf{E}_{1}}} \min _{\gamma_{i j}} c\left(\mathbf{x}_{i}, \mathbf{x}_{j}\right)$ is the sum of the minimum cost from each terminal that is a  neighbour of  $\mathbf{x}_{j}$ to $\mathbf{x}_{j}$ itself,  which can be calculated using FMM with the terminal being the source node. Let $\bar{c}\left(\mathbf{x}_{j}\right)=0$ if no terminal is adjacent to the Steiner node $\mathbf{x}_{j}$.

\begin{figure}[htbp]
    \centering
    \includegraphics[width = 8cm]{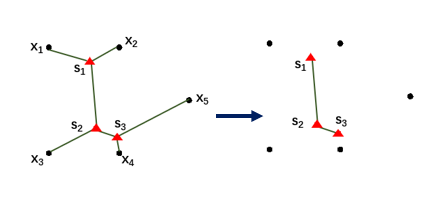}
    \caption{A Steiner tree and its skeleton tree.}
    \label{fig:skeleton}
\end{figure}

We define the \emph{skeleton tree} of the graph $\mathcal{T}$  to be  the subtree  $\mathbb{T}=\left(\mathbf{S}, \mathbf{E}_{2}\right)$  composed of only Steiner nodes and the edges connect them to each other. This is illustrated  in Fig.~\ref{fig:skeleton}.

\begin{figure}[htbp]
	\centering
	\includegraphics[width =3.5cm]{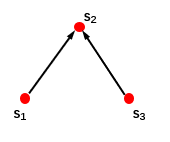}
	\caption{The skeleton tree with $\mathbf{s}_2$ being the root in Fig.~\ref{fig:skeleton}. One reordered  Steiner node sequence is $\mathbf{s}_1$, $\mathbf{s}_3$, $\mathbf{s}_2$.}
	\label{fig:directed}
\end{figure}

 For a tree with a Steiner topology $\mathcal{T}$; that is,  with each Steiner node  having three branches and each terminal node three or less branches, we choose an arbitrary Steiner node $\mathbf{s}_r$ as the root of its skeleton tree $\mathbb{T}$, $\mathbf{s}_r \in \mathbb{T}$. Next, we assign to the edges of $\mathbb{T}$ an orientation towards the root. Then, $\mathbb{T}$ becomes a directed rooted tree (i.e., anti-arborescence). See Fig.~\ref{fig:directed}. For more details on directed rooted trees, the reader is referred to~\cite{wang2020optimal}. We order the nodes of the skeleton tree so that children  of a given node  $\mathbf{s}_i$ appear in the list earlier than the given node~(that is, with arrows \emph{from} them to $\mathbf{s}_i$). An order with this property is referred to as a \emph{topological order}.
We denote, without loss of generality, the reordered Steiner node sequence as $1, 2, \ldots, M$, where $M$ corresponds to the root of $\mathbb{T}$, and denote $\overline{\mathbf{x}_{i}}=\mathbf{x}_{N+i}, i=1,2,...,M.$  Altering the order of Steiner nodes $\mathbf{x}_{N+i} \in \mathbb{M}$ in Eq.~(\ref{equ:mincostofnetwork}) does not change the optimization problem, so we can rewrite Eq.~(\ref{equ:mincostofnetwork}) as Eq.~(\ref{equ:mincostofnetwork2}).
\begin{equation}
\label{equ:mincostofnetwork2}
\begin{aligned}
&\min\limits _{X}\!\Psi(\mathbf{X}, \Gamma) =\min\limits _{\overline{\mathbf{x}}_{M}, \overline{\mathbf{x}}_{M-1}, \ldots, \overline{\mathbf{x}}_{1}}\Phi\left(\overline{\mathbf{x}}_{M}, \overline{\mathbf{x}}_{M-1}, \ldots, \overline{\mathbf{x}}_{1}\right)\\
\end{aligned}\end{equation}
where
\begin{equation}
\label{equ:mincostofnetwork22}
\begin{aligned}
&\Phi\left(\overline{\mathbf{x}}_{M}, \overline{\mathbf{x}}_{M-1}, \ldots, \overline{\mathbf{x}}_{1}\right)\\
&=\bar{c}\left(\overline{\mathbf{x}}_{M}\right)+\!\!\!\!\sum_{(j, M) \in \mathbf{E}_{2} \atop j \in C(M)} \min _{\gamma_{j M}} c\left(\overline{\mathbf{x}}_{j}, \overline{\mathbf{x}}_{M}\right)+\Phi\left(\overline{\mathbf{x}}_{M-1}, \ldots, \overline{\mathbf{x}}_{1}\right)\\
\end{aligned}\end{equation}
where $C(M)$ is the children set of node $M$.

We define $\Phi(\overline{\mathbf{x}}_{i})=\bar{c}(\overline{\mathbf{x}}_{i})$ for any leaf $i$, and
\begin{equation}
\begin{aligned}
&\Phi^{*}\left(\overline{\mathbf{x}}_{M}, \overline{\mathbf{x}}_{M-1}, \ldots, \overline{\mathbf{x}}_{1}\right)\\
&=\min _{\bar{\mathbf{x}}_{M},\bar{\mathbf{x}}_{M-1},...,\bar{\mathbf{x}}_{1}} \Phi\left(\overline{\mathbf{x}}_{M}, \overline{\mathbf{x}}_{M-1}, \ldots, \overline{\mathbf{x}}_{1}\right)
\end{aligned}
\end{equation}

Then, the problem shown in Eq.~(\ref{equ:mincostofnetwork}) is converted to a dynamic programming problem with Bellman equation:
\begin{multline}
\label{equ:Bellman}
\Phi^{*}(\overline{\mathbf{x}}_{M}, \overline{\mathbf{x}}_{M-1}, \ldots, \overline{\mathbf{x}}_{1}) \\
=\min_{\overline{\mathbf{x}}_{M}}\Bigl[\bar{c}(\overline{\mathbf{x}}_{M})+\sum_{(j, M) \in E_{2} \atop j \in C(M)} \min _{\gamma_{j M}} c(\overline{\mathbf{x}}_{j}, \overline{\mathbf{x}}_{M})\\
\qquad\qquad  +\Phi^{*}(\overline{\mathbf{x}}_{M-1}, \ldots,\overline{\mathbf{x}}_{1})\Bigr]. 
\end{multline}

\begin{figure}[htbp]
	\centering
	\includegraphics[width =5cm]{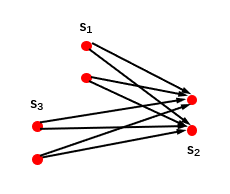}
	\caption{The DAG corresponding to the Steiner topology and skeleton tree in Fig.~\ref{fig:skeleton}.}
	\label{fig:directed2}
\end{figure}

To solve this dynamic programming problem, we construct a new directed acyclic graph (DAG) $\mathbf{G}=$ $\left(\mathbf{V}^{\prime}, \mathbf{E}^{\prime}\right)$ based on $\mathbb{T}$, as shown in Fig.~\ref{fig:directed2}. Each Steiner node $i \in \mathbb{T}$ is associated with a subset $\mathbf{A}_{i}$ of $\mathbf{V}^{\prime},$ where $\mathbf{A}_{i}$ are grid nodes of $\mathbb{M}$ and the weight on each node is $\bar{c}_{i}(\mathbf{x})$. It follows that $\mathbf{V}^{\prime}=\cup_{i \in \mathbf{S}} \mathbf{A}_{i}$; that is, $\mathbf{V}^{\prime}$ is composed of $m$ copies of the grid nodes of $\mathbb{M}$. For an arc $e=(i, j) \in \mathbf{E}_{2}$, where $\mathbf{s}_j$ is the parent of $ \mathbf{s}_i$, we construct a \emph{complete connection} from $\mathbf{A}_{i}$ to $\mathbf{A}_{j}$ for $\mathbf{G}$; that is, an arc $\varepsilon=(p,q)$ from every $\mathbf{x}_p \in \mathbf{A}_{i}$ to every $\mathbf{x}_q \in \mathbf{A}_{j}$ in $\mathbf{G} $.  The cost of the arc $\varepsilon$ is defined as the minimum cost from node $\mathbf{x}_{p}$ to node $\mathbf{x}_{q},$ calculated by FMM, i.e., $w(\varepsilon)=w(p,q)=\min c\left(\mathbf{x}_{p}, \mathbf{x}_{q}\right)$. We define $\phi^{i}_{p}$ as the  minimum cumulative cost (MCC) for a node $\mathbf{x}_{p} \in \mathbf{A}_{i}$: 
\begin{equation}
\label{equ:dynamic}
   \phi_{p}^{i}=\bar{c}\left(\mathbf{x}_{p}\right)+\sum_{j \in C(i)} \min _{q \in \mathbf{A}_{j}}\left(w\left(\mathbf{x}_{q}, \mathbf{x}_{p}\right)+\phi_{q}^{j}\right).
\end{equation}

For the problem of Steiner tree total cost optimization taking into account  the cost of Steiner nodes (BUs in the cable system), we have  a new Bellman equation as shown in Eq.~(\ref{equ:BellmanBU}), which is an  extension of Eq.~(\ref{equ:Bellman}). 
\newcounter{subsec:SMT_BU}
\setcounter{subsec:SMT_BU}{\value{equation}} 
\setcounter{equation}{7} 
\begin{figure*}[] 
	\hrulefill
\begin{multline}
\label{equ:BellmanBU}
\Phi_{BU}^*(\overline{\mathbf{x}}_{M}, \overline{\mathbf{x}}_{M-1}, \ldots, \overline{\mathbf{x}}_{1})
=\min_{\overline{\mathbf{x}}_{M}}\Bigl[\Phi_{BU}^*(\overline{\mathbf{x}}_{M-1}, \ldots, \overline{\mathbf{x}}_{1})+\bigl(\bar{c}(\overline{\mathbf{x}}_{M})+\mathbf{b}_M\bigr)
+\sum_{(j, M) \in \mathbf{E}_{2} \atop j \in C(M)} \min _{\gamma_{j M}} c\left(\overline{\mathbf{x}}_{j}, \overline{\mathbf{x}}_{M}\right)\Bigr]
\end{multline}
\end{figure*}

We refer to the DAG-Least-Cost-Tree algorithm in~\cite{wang2020optimal} that finds the tree on DAG with the minimum cost and return the coordinates of the Steiner nodes. We propose a new algorithm --- the DAG-Least-Cost-System algorithm --- listed as Algorithm~\ref{alg:BUcost} that  finds the tree on DAG with the minimum total cost (length and number of Steiner nodes) and return the coordinates of the Steiner nodes. The extension of the DAG-Least-Cost-System algorithm over our earlier  DAG-Least-Cost-Tree~\cite{wang2020optimal} enables the inclusion of  weights (costs) of the Steiner nodes. This extension has to take account of   Steiner nodes with degree in excess of three. The extension can also be used to optimize connections between regions, as opposed to points, as we will demonstrate here. 

\begin{algorithm}[htbp] 
\caption{DAG-Least-Cost-System} 
\label{alg:BUcost}
\begin{algorithmic}[1] 
\REQUIRE~~\\ 
 The graph $\mathbf{G}=(\mathbf{V}',\mathbf{E}')$, $\mathbb{T}$ with  a topological order and  BU cost ($\mathbf{b}_i$) for each mesh point.
\ENSURE ~~\\ 
 Coordinates of Steiner nodes  $\mathbf{s}_{i}, i=1, \ldots, M$.
\FOR{$i=1,...,M$}
\FOR{each node $\mathbf{x}_{p} \in \mathbf{A}_{i}$}
\IF{$i$ is a leaf in $\mathbb{T}$}
\STATE $\phi_{p}^{i}=\bar{c}\left(\mathbf{x}_{p}\right)+\mathbf{b}_p ; \pi\left(\mathbf{x}_{p}\right)=\mathbf{NIL}$;
\ELSE
\FOR{each child $\mathbf{s}_j$ of $\mathbf{s}_i$}
\STATE $\pi\left(\mathbf{x}_{p}, j\right)=\mathrm{NIL}$;
\ENDFOR
\STATE $\phi_{p}^{i}=0$;
\ENDIF
\ENDFOR
\ENDFOR
\FOR{$i=1,...,M$, $i$ is not a leaf}
\FOR{each node $\mathbf{x}_{p} \in \mathbf{A}_{i}$}
\FOR{each child $s_j$ of $s_i$}
\STATE $\psi=\infty$;
\FOR{each node $\mathbf{x}_{q} \in \mathbf{A}_{j}$}

\IF{$\mathbf{x}_q=\mathbf{x}_p$}
\STATE $\psi'=\phi_{q}^{j}+w\left(\mathbf{x}_{q}, \mathbf{x}_{p}\right)$
\ELSE
\STATE $\psi'=\phi_{q}^{j}+w\left(\mathbf{x}_{q}, \mathbf{x}_{p}\right)+\mathbf{b}_q$
\ENDIF
\IF{$\psi>\psi'$}
\STATE $\psi=\psi'$;\\
$\pi\left(\mathbf{x}_{p}, j\right)=q;$
\ENDIF
\ENDFOR
\STATE $\phi_{p}^{i}=\phi_{p}^{i}+\psi$;
\ENDFOR
\STATE $\phi_{p}^{i}=\phi_{p}^{i}+\bar{c}\left(\mathbf{x}_{p}\right)$;
\ENDFOR
\ENDFOR
\STATE Let $\hat{p}_{M}=\arg \min _{\mathbf{x}_{p} \in \mathbf{A}_{M}} \phi_{p}^{M}$
\STATE Trace back from $\hat{p}_{M}$ to leaves via $\pi$;
\RETURN $\mathbf{x}_{\hat{p}_{1}}, \ldots, \mathbf{x}_{\hat{p}_{M}}$.
\end{algorithmic}
\end{algorithm}

In Algorithm \ref{alg:BUcost}, the statements in Lines $1$-$12$ form the initialization. Implementation of Eq.~ (\ref{equ:BellmanBU}) is in Lines $13$-$32$. Note that $\mathbb{T}$ is the skeleton of a Steiner tree with full Steiner topology. If a leaf node in $\mathbb{T}$ has the same coordinates as a terminal node then the BU cost $\mathbf{b}_i$ at this location is zero. If two Steiner nodes have the same location, we need to decide whether to count the cost of the Steiner node once or twice depending on the BU branch constraints (Lines $18$-$22$ achieve this).  As in~\cite{wang2020optimal}, once the iterations reach the root, in Line~$32$, we choose the grid node $\hat{\mathbf{p}}$ (with MCC $\phi_{p}^{M}$) in $\mathbf{A}_{M}$. The node $\hat{\mathbf{p}}$ is the physical Steiner node corresponding to the root. To derive the coordinates of the remaining Steiner nodes, we track back on $\mathbf{G}$ starting from $\hat{\mathbf{p}}$. 
From the locations of the Steiner nodes, we  infer the number of BUs (Steiner nodes). 

\subsection{Minimal Cost Steiner Tree for  Regional Connectivity}
\label{subsec:SMT_area}
 We aim, in this subsection, to find a minimum total cost network connecting multiple regions and terminal nodes. To solve the regional connectivity problem, firstly  we  run FMM from each of the nodes in each  region to every grid node in $\mathbb{M}$ to derive $\bar{c}\left(\mathbf{x}_{j}\right)$: the sum of the minimum cost from each region that  is a neighbour of  $\mathbf{x}_{j}$ to $\mathbf{x}_{j}$ itself. Unlike in  the terminal node connection problem, FMM needs to be applied for every nodes in the region. However, for a submarine cable system, there are just a few locations on the coastline that need to be considered. In fact, it is clear that minimal lengths are achieved by connecting to nodes on the coastline (or at least on the boundaries of the regions). For each of these nodes, $\mathbf{r}_{il}$, in the region $\mathbf{R}_{i}$, we calculate the distance $\mathbf{d}_{il}(\mathbf{x}_{j})$, where $l$ indicates the index of a possible location on the boundary of  $\mathbf{R}_i$. The minimal cost from  each  such node in this region to each grid node is written as $\mathbf{D}_i\left(\mathbf{x}_{j}\right)$. We define  the  ``pointer matrix",  $\mathbf{P}_{ij}$, to  indicate the location (the value of $l$) in region $\mathbf{R}_i$ corresponding to $\mathbf{D}_i\left(\mathbf{x}_{j}\right)$.

 Algorithm~\ref{alg:BUcost} summarises the solution for minimizing the total cost  for the  terminal node connectivity problem whereas the procedure we summarized below is for minimizing total cost  in the regional connectivity problem.

\begin{enumerate}
\item For each region $\mathbf{R}_i$, run FMM (region to points), and calculate the distance record $\mathbf{D}_i(\mathbf{x}_{j})$ and a corresponding pointer matrix $\mathbf{P}_{ij}$.

\item For each Steiner node $\mathbf{s}_i, i \in \mathbf{S},$ let $\bar{c}_{i}\left(\mathbf{x}_{j}\right)=$ $\sum_{l \in \mathbf{N},(l,i) \in \mathbf{E}_{1}} \mathbf{D}_l\left(\mathbf{x}_{j}\right)$ for each grid node $\mathbf{x}_{j} \in \mathbb{M}$, where $\mathbf{E}_1$ is based on the given topology $\mathcal{T}$;

\item For each pair of grid nodes $\mathbf{x}_{i}, \mathbf{x}_{j} \in \mathbb{M},$ run FMM for calculating the minimum cost from $\mathbf{x}_{i}$ to $\mathbf{x}_{j}$;

\item Based on the Steiner topology $\mathcal{T}$ and the grid nodes of $\mathbb{M},$ construct the $\mathrm{DAG} ~\mathbf{G}=\left(\mathbf{V}^{\prime}, \mathbf{E}^{\prime}\right)$;

\item Run Algorithm~\ref{alg:BUcost} on the $\mathrm{DAG}$ and find the minimum cost network. The nodes on the minimum cost network $\mathbf{x}_{\hat{p}_{1}}, \ldots, \mathbf{x}_{\hat{p}_{M}}$, are the Steiner nodes;

\item Find the geodesics $\Gamma=\{\gamma(e) \mid e \in \mathbf{E}\}$ by gradient  descent, as the last step in  FMM,  and the chosen location in each region taking into account the corresponding pointer matrix $\mathbf{P}_{ij}$.
\end{enumerate}

\subsection{Computational Complexity Analysis}
We assume $\mathbb{M}$ consists of $H$ grid nodes and $N$  terminals/regions to be connected. For the regional connectivity problem, without loss of generality, we assume every region has $L$ nodes. In a similar way it is done in~\cite{wang2020optimal}, the complexity of step $1$ in the procedure provided above is $\mathcal{O}((2L-1) NH \log H)$,  and the complexity of step $2$ is $\mathcal{O}((N-2) H)$ (for a full Steiner topology with $N-2$ Steiner nodes). Step 3 requires the cost calculation of each pair of grid nodes in $\mathbb{M}$, and its complexity is $\mathcal{O}\left(H^{2} \log H\right)$. The DAG  has $(N-2)H$ vertices and $(N-3) H^{2}$ arcs. Therefore, finding the minimal cost tree on $G$ takes $\mathcal{O}\left((N-2)H+(N-3) H^{2}\right)$ operations. Therefore, the computational complexity of the whole algorithm is $\mathcal{O}\left(H^{2}(\log H+N-3)\right)$.

\section{Applications}\label{sec:num_res}
\begin{figure}[htbp]
	\centering
	\includegraphics[width =0.9\columnwidth]{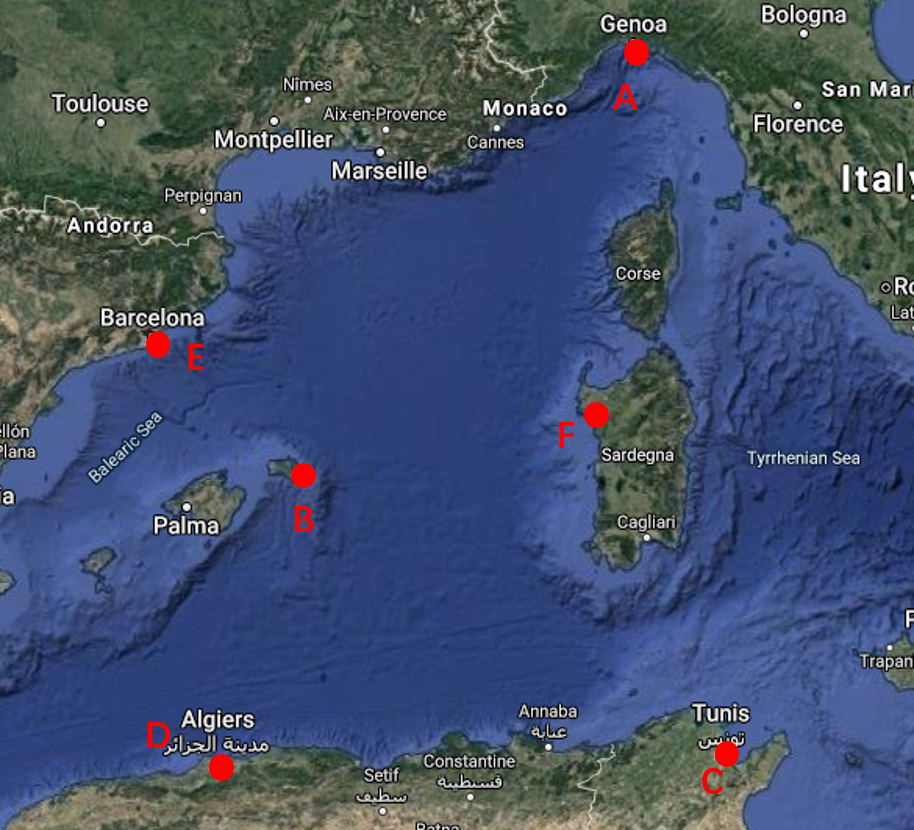}
	\caption{Region $\mathbb{D}$. Source: Google Earth.}
	\label{fig:area}
\end{figure}

In this section, we present an application of our method to several realistic scenarios. As in~\cite{wang2020optimal}, we use bathymetric data from the Global Multi-Resolution Topography synthesis~\cite{GMRT}. The object region $\mathbb{D}$ spans from the northwest corner ($45.000\degree \text{N}$, $0.000\degree \text{E}$) to the southeast corner ($36.000\degree \text{N}$, $11.000\degree \text{E}$). See Fig.~\ref{fig:area}.
All the examples in this section take place in this region. In our examples, we demonstrate how to choose the location and number of BUs and the location choices of CLSs for minimization of the total cost of a cable system.  

\subsection{The Effect of BU Cost on Cable Network Design}
\label{subsec:BUcosteffect}

In this subsection, we plan a submarine cable network using a trunk-and-branch topology between the following five locations: Genoa ($44.407\degree \text{N}$, $8.963\degree \text{E}$), Palma ($32.828\degree \text{N}$, $4.310\degree \text{E}$), Tunis ($37.341\degree \text{N}$, $9.078\degree \text{E}$),  Algiers ($36.832\degree \text{N}$, $3.052\degree \text{E}$),  Algiers ($36.832\degree \text{N}$, $3.052\degree \text{E}$) and Barcelona  ($41.399\degree \text{N}$, $2.085\degree \text{E}$); these are denoted as A, B, C, D, E, F, respectively.

As discussed in Introduction, most existing BUs have three branches. We  make this assumption in  most of our experiments,  consistent with the characteristics of the Steiner tree; in Section~\ref{subsec:BUbranchmul} we deal with multi-branch BUs. 



As the Steiner nodes represent BUs, an increase in the number of BUs (Steiner nodes) may adversely affect the total network cost. Therefore, our optimization must consider the tradeoff between the cost of BUs against the benefit in reduction of total cable length. 
 
\begin{figure}[htbp]
\centering
\subfigure[Cable system with two BUs.]{
\includegraphics[width=5cm]{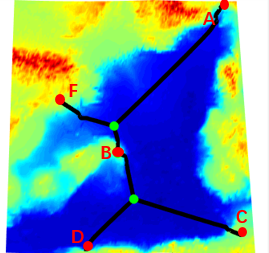}
\label{fig:bucost2}
}
\subfigure[Cable system with one BU.]{
\includegraphics[width=5cm]{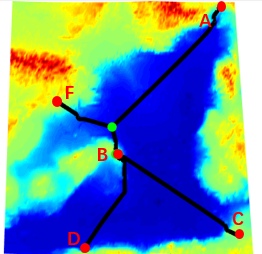}
\label{fig:bucost1}
}

\subfigure[Minimum Spanning Tree.]{
\includegraphics[width=5cm]{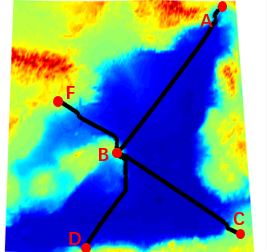}
\label{fig:bucost0}
}
\caption{Cable network result changed by the BU cost.}
\label{fig:bucostall}
\end{figure}

To be realistic, as in~\cite{wang2021optimal,wang2019cost,Elias}, we assume here that the cost of submarine cable construction is around $\$$25,000 per kilometer and the cost of a CLS is $\$10$ million. The cost of BUs are assumed to vary between $\$1$-$3$ million. Notice that this range is somewhat wider than the industry estimate of $\$1$-$2$ million~\cite{Paul,Dick}. We will observe, as expected,  that an increase in  BU cost leads to a decrease in their quantity. Figs.~\ref{fig:bucost2}-\ref{fig:bucost0} show the result of our method.

Table~\ref{tab:BUCOST} shows the details  of each cable network with varying cost of BUs. From the result, we can conclude that, with  increasing  installation cost of BU, the number of BUs in the cable network decreases, while  the total length of cable network becomes larger. Also as expected, the total cost of the network also rises with the increasing of BU cost, but the total cost is always the optimal solution under the specified BU cost.

\begin{table}[t]
\renewcommand{\arraystretch}{1.3}
\caption{Effect of BU cost on optimal cable network design.}
\label{tab:BUCOST}
\centering
\begin{tabular}{|c|c|c|c|}
\hline BU cost (million \$)  & $1$  & $2$  & $3$ \\
\hline Optimal number of BUs  & $2$ & $1$  &  $0$\\
\hline Total length~(km) &1609.04& 1777.95 & 1822.05 \\
\hline Total cost~(million \$) & 92.22 & 95.45 & 95.55
\\
\hline Cable network & Fig.~\ref{fig:bucost2} & Fig.~\ref{fig:bucost1} & Fig.~\ref{fig:bucost0} \\
\hline
\end{tabular}
\end{table}

\begin{figure}[htbp]
	\centering
	\includegraphics[width = 0.7\columnwidth]{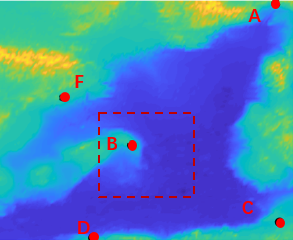}
	\caption{Region $\mathbb{D}$ with spacial area.}
	\label{fig:area2}
\end{figure}

\begin{figure}[htbp]
	\centering
	\includegraphics[width = 0.7\columnwidth]{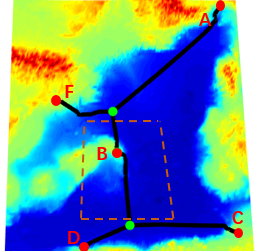}
	\caption{Cable network for Region $\mathbb{D}$ with a special higher cost area.}
	\label{fig:BUCOST4}
\end{figure}

As we discuss in Section~\ref{sec:Introduction}, seabed conditions influence installation costs of BUs:  BU installation cost in some areas is relatively higher than others. To illustrate the effects of geographically varying BU cost, we redo the previous experiment with an assumed higher installation costs for BUs of \$2 million across the area from northwest corner ($40.696\degree \text{N}$, $2.710\degree \text{E}$) to southeast corner ($38.375\degree \text{N}$, $5.478\degree \text{E}$), as shown in Fig.~\ref{fig:area2}. The remainder of the area has a BU installation cost of  \$1 million. The results are shown in  Fig.~\ref{fig:BUCOST4}, where we can see that the locations of the two BUs tend to avoid the higher cost area. 
The total cost of the derived network is $\$94.88$ million with the total length being $1715.32$~km.

\subsection{Multi-branch BUs With More Than Three Branches}
\label{subsec:BUbranchmul}

As discussed, our algorithm is based on the Steiner tree algorithm used in \cite{wang2020optimal} which considers all full Steiner topologies and applies  Smith's $\mathrm{B} \& \mathrm{B}$ method to reduce the computation. 

Here we use the extension of this algorithm  described in Section~\ref{sec:method},  based on Theorem~1,  to the case where the cost  of  BUs is taken into account,  to provide a realistic example where the optimal solution involves a BU with more than three branches.

Specifically, we consider an example where our aim is to connect the following four locations,  Palma ($32.828\degree \text{N}$, $4.310\degree \text{E}$), Tunis ($37.341\degree \text{N}$, $9.078\degree \text{E}$),  Algiers ($36.832\degree \text{N}$, $3.052\degree \text{E}$) and Sardegna ($40.568\degree \text{N}$, $8.333\degree \text{E}$), as shown in Fig.~\ref{fig:area} and  denoted by B, C, D, F, respectively.

\begin{figure}[htbp]
\centering
\subfigure[BUs with more than three branches.]{
\includegraphics[width=5cm]{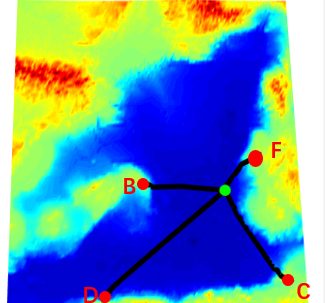}
\label{fig:BUmul1}
}
\subfigure[BUs with only three branches.]{
\includegraphics[width=5cm]{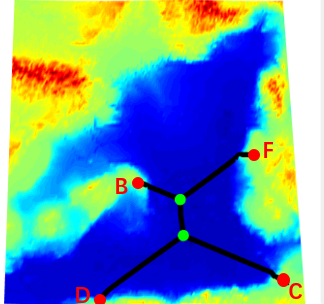}
\label{fig:BUmul2}
}
\caption{Cable network optimization results with different degrees of BU.}
\label{fig:bucostall}
\end{figure}

First, we assume that the cost of a BU is \$1.7 million across the entire area, the average cable cost is \$25,000, and the number of branches of BUs is not limited.  This results in an optimal tree that has just one BU with four branches and  the total length is $1190.71$~km, as Fig.~\ref{fig:BUmul1} shows. Then, to illustrate the benefit of multi-branch BUs, we present in  Fig.~\ref{fig:BUmul2} the result of the algorithm for the situation where we restrict BUs to have only three branches and apply our algorithm while preventing  combining of Steiner nodes. The result shows that its total cable length is $1127.86$~km which is smaller than the result in Fig.~\ref{fig:BUmul1}. However, the cost of the cable network in Fig.~\ref{fig:BUmul1}, namely, \$31,467,834.95 is less than that of the solution presented in Fig.~\ref{fig:BUmul2} which is \$31,596,702.04.  We remind the reader that the wide majority of cable systems have only 3-branch BUs. Nevertheless, we have demonstrated an example where it is beneficial to use 4-branch BUs. 

\subsection{SMT for Regions Connecting}
\label{subsec:regionsimu}
In this section, we give  an example to illustrate our regional connectivity solution for optimizing the location of a CLS in each region to minimize total cost. To this end,  we again  apply our algorithm to the  area shown in Fig.~\ref{fig:area} to show the benefit of optimizing the CLS locations in the relevant regions. We assume a constant BU cost of \$1 million which is same as that used in the example illustrated in Fig.~\ref{fig:bucost2}. 

\begin{figure}[htbp]
	\centering
	\includegraphics[width = 0.7\columnwidth]{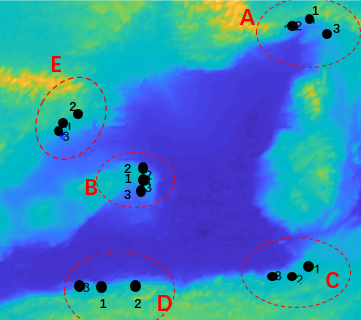}
	\caption{Regional connectivity problem description.}
	\label{fig:areaconnect}
\end{figure}

\begin{figure}[htbp]
	\centering
	\includegraphics[width = 0.7\columnwidth]{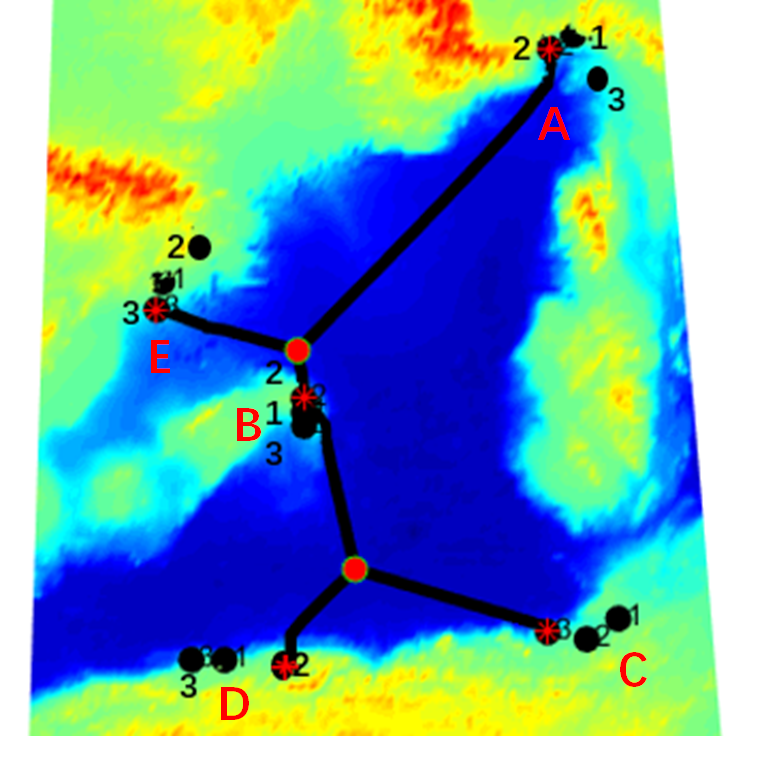}
	\caption{Multi-CLSs with same cost in each region.}
	\label{fig:areacon1}
\end{figure}

We consider five regions corresponding to the five locations A, B, C, D, E. That is, the first region is in the vicinity of location A, the second  in the vicinity of location B, etc. In each region, we assume there are three potential CLS locations. One of the potential CLS locations in each region is the same as the original locations A, B, C, D, E, and the other two are arbitrarily set, as shown in Fig.~\ref{fig:areaconnect}.  Accordingly, there are three alternative CLS locations in each region, and we specify the original nodes as the first among the three alternatives. First, we assume that the construction costs of CLSs in all these fifteen locations are the same, and we set each CLS cost as  $\$10$ million. The result of our optimization algorithm are presented in Fig.~\ref{fig:areacon1}. In regions A, B, D, the second locations are chosen as the CLS locations, while in regions C and E, the third locations are chosen.  The total length of the optimal cable network is $1512.56$~km, which is lower by approximately $6\%$ than if we just choose the original locations for the CLSs as shown in Fig.~\ref{fig:bucost2} in Section~\ref{subsec:BUcosteffect}. Further details of this cable network optimization are shown in Table~\ref{tab:lineCOST2}.

\begin{table}[t]
\renewcommand{\arraystretch}{1.3}
\caption{Multi-CLSs with same cost in each region.}
\label{tab:lineCOST2}
\centering
\begin{tabular}{|c|c|c|c|c|c|}
\hline  Region & $A$ & $B$  & $C$& $D$& $E$ \\
\hline Station choice & (2) & (2) & (3) & (2) & (3)\\
\hline Total length~(km) & \multicolumn{5}{|c|}{1512.56}\\
\hline Total cost~(million \$) & \multicolumn{5}{|c|}{89.81}\\
\hline 
\end{tabular}
\end{table}

\begin{figure}[htbp]
	\centering
	\includegraphics[width = 0.7\columnwidth]{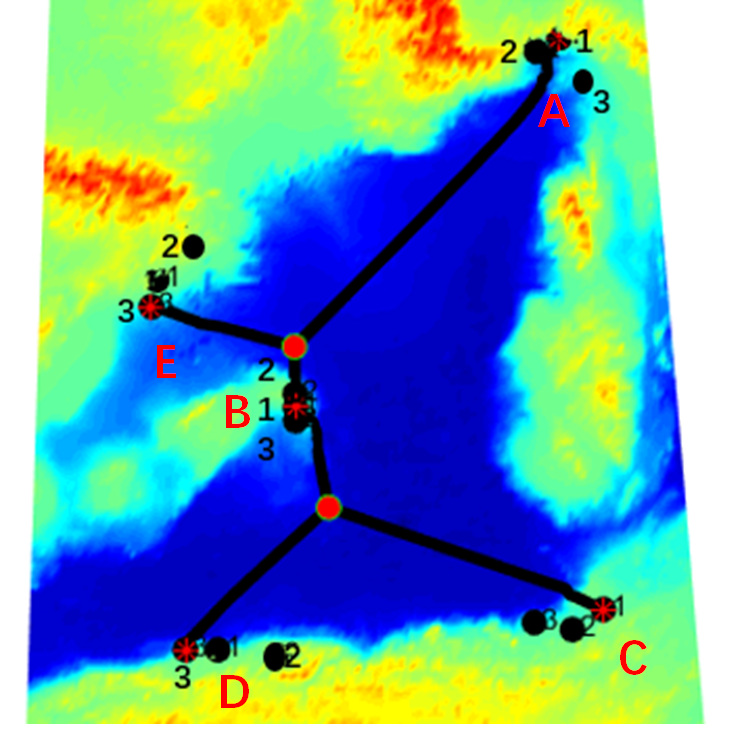}
	\caption{Multi-CLSs with different cost in each region.}
	\label{fig:areacon2}
\end{figure}

As discussed in Introduction, the construction costs of CLSs is determined by many factors and may vary from region to region. In the next example, we study the situation where different CLSs have different cost, and use our method to find the optimal solution. The construction cost of each CLS is shown in Table~\ref{tab:lineCOST1}. The optimal cable network results are shown in Fig.~\ref{fig:areacon2} and the details are shown in Table~\ref{tab:lineCOST1}. 

\begin{table}[t]
\renewcommand{\arraystretch}{1.3}
\caption{Multi-CLSs with different cost in each region.}
\label{tab:lineCOST1}
\centering
\begin{tabular}{|c|c|c|c|c|c|}
\hline  Region  & $A$ & $B$  & $C$& $D$& $E$ \\
\hline First Station cost (million \$)  & 10 & 10 & 10& 15& 15\\
\hline Second Station cost (million \$) &15& 14 & 15& 16& 14\\
\hline Third Station cost (million \$) &13& 12 & 14& 10& 10\\
\hline Station choice &(1)&(1)&(1)&(3)&(3)\\
\hline Total length~(km) & \multicolumn{5}{|c|}{1634.20}\\
\hline Total cost~(million \$) &\multicolumn{5}{|c|}{92.85}\\
\hline 
\end{tabular}
\end{table}

The result in Fig.~\ref{fig:areacon2} shows that the CLS locations with lower construction cost will be chosen to minimize the total cost though the total length is $8\%$ longer than the result in Fig.~\ref{fig:areacon1}. In regions A, B, C, the first locations are chosen as the CLS locations, while in regions B and E, the third locations are chosen.

\section{Conclusion}
\label{sec:conclusion}
We have articulated  a solution to the problem of submarine cable network path planning where the topology is trunk-and-branch network. This is done for a network  on the surface of the earth and takes into consideration  the cost of BUs, CLSs and cable laying as well as the selection of locations  of CLSs.  

We have introduced the weighted Steiner node problem, which in mathematical terms, has been couched as a variant of the SMT problem, on an irregular 2-dimensional manifold in $\mathbb{R}^{3}$. The resulting algorithm has polynomial-time computational complexity. The Steiner nodes in our problem can vary in number, while incurring a penalty (cost), and their degrees are not constrained to three. We have proposed an algorithm that can solve this variant of the Steiner tree problem and have proved that our algorithm is applicable to Steiner nodes with degrees of three or more, so that it can be used to optimize the total cost of a cable network taking account of  the costs of BUs and CLSs.

We have applied our technique  to several realistic scenarios to elucidate that our method can achieve  an optimal solution for realistic cable network optimization problems. Our technique can also be applied to other network design problems on irregular 2-dimensional manifolds,  aside from the  submarine cables. These include problems involving  power distribution and  gas and oil pipelines.


\begin{thebibliography}{10}
\providecommand{\url}[1]{#1}
\csname url@samestyle\endcsname
\providecommand{\newblock}{\relax}
\providecommand{\bibinfo}[2]{#2}
\providecommand{\BIBentrySTDinterwordspacing}{\spaceskip=0pt\relax}
\providecommand{\BIBentryALTinterwordstretchfactor}{4}
\providecommand{\BIBentryALTinterwordspacing}{\spaceskip=\fontdimen2\font plus
\BIBentryALTinterwordstretchfactor\fontdimen3\font minus
  \fontdimen4\font\relax}
\providecommand{\BIBforeignlanguage}[2]{{%
\expandafter\ifx\csname l@#1\endcsname\relax
\typeout{** WARNING: IEEEtran.bst: No hyphenation pattern has been}%
\typeout{** loaded for the language `#1'. Using the pattern for}%
\typeout{** the default language instead.}%
\else
\language=\csname l@#1\endcsname
\fi
#2}}
\providecommand{\BIBdecl}{\relax}
\BIBdecl

\bibitem{Submarinereport2020}
\BIBentryALTinterwordspacing
{A. McCurdy \textit{et al.}}, ``Submarine telecoms industry report,'' Submarine
  Telecoms Forum, Inc., Sterling, Virginia, USA, Tech. Rep., October
  2020~(accessed on March 10, 2021). [Online]. Available:
  \url{https://subtelforum.com/products/submarine-telecoms-industry-report/}
\BIBentrySTDinterwordspacing

\bibitem{cablequestions}
{TeleGeography}, ``Submarine cable frequently asked questions,''
  \url{https://www2.telegeography.com/submarine-cable-faqs-frequently-asked-questions},
  2021~(accessed on March,2021).

\bibitem{Basics}
\BIBentryALTinterwordspacing
A.~Markow, ``{Summary of Undersea Fiber Optic Network Technology and
  Systems},'' October 2017. [Online]. Available:
  \url{{http://hmorell.com/subcable/documents/basics}}
\BIBentrySTDinterwordspacing

\bibitem{Offshore}
\BIBentryALTinterwordspacing
``{Offshore Renewable and Cable Awareness},'' October 2019. [Online].
  Available: \url{https://kis-orca.eu/{subsea-cables}/design/}
\BIBentrySTDinterwordspacing

\bibitem{Paul}
\BIBentryALTinterwordspacing
``{Responses to the two follow-up questions directed to Paul Shorb},''
  September 2002. [Online]. Available:
  \url{http://sdb0947dd74429a3f.jimcontent.com/\\
  download/version/1391619811/module/14390269-\\ 13/
  name/ResponsestoOceansCommissionQuest-\\ ions.pdf}
\BIBentrySTDinterwordspacing

\bibitem{Dick}
D.~Hale, personal communication, March 2021.

\bibitem{Elias}
E.~Tahchi, personal communication, Oct. 2016.

\bibitem{ye2018submarine}
Y.~YE, X.~Jiang, G.~Pan, and W.~Jiang, \emph{Submarine Optical Cable
  Engineering}.\hskip 1em plus 0.5em minus 0.4em\relax Academic Press, 2018.

\bibitem{SUBMARINECABLENETWORK}
\BIBentryALTinterwordspacing
W.~Qiu, ``{Submarine Cables Crossing Egypt and their costs},'' April 2020.
  [Online]. Available: \url{https://www.submarinenetworks.com/en/services/\\
  research/submarine-cables-crossing-egypt-and-cost}
\BIBentrySTDinterwordspacing

\bibitem{wang2021jlt1}
X.~Wang, Z.~Wang, E.~Tahchi, and M.~Zukerman, ``Submarine cable path
  optimization based on weight selection of design considerations,'' Submitted
  for publication.

\bibitem{wang2019path}
Q.~Wang, Z.~Wang, J.~Guo, E.~Tahchi, X.~Wang, B.~Moran, and M.~Zukerman, ``Path
  planning of submarine cables,'' in \emph{2019 21st International Conference
  on Transparent Optical Networks (ICTON)}.\hskip 1em plus 0.5em minus
  0.4em\relax IEEE, 2019, pp. 1--4.

\bibitem{wang2017multiobjective}
Z.~Wang, Q.~Wang, M.~Zukerman, J.~Guo, Y.~Wang, G.~Wang, J.~Yang, and B.~Moran,
  ``Multiobjective path optimization for critical infrastructure links with
  consideration to seismic resilience,'' \emph{Computer-Aided Civil and
  Infrastructure Engineering}, vol.~32, no.~10, pp. 836--855, 2017.

\bibitem{zhao2016route}
M.~Zhao, T.~W. Chow, P.~Tang, Z.~Wang, J.~Guo, and M.~Zukerman, ``Route
  selection for cabling considering cost minimization and earthquake
  survivability via a semi-supervised probabilistic model,'' \emph{IEEE
  Transactions on Industrial Informatics}, vol.~13, no.~2, pp. 502--511, 2016.

\bibitem{wang2019cost}
Q.~Wang, J.~Guo, Z.~Wang, E.~Tahchi, X.~Wang, B.~Moran, and M.~Zukerman,
  ``Cost-effective path planning for submarine cable network extension,''
  \emph{IEEE Access}, vol.~7, pp. 61\,883--61\,895, 2019.

\bibitem{wang2020optimal}
Z.~Wang, Q.~Wang, B.~Moran, and M.~Zukerman, ``Optimal submarine cable path
  planning and trunk-and-branch tree network topology design,'' \emph{IEEE/ACM
  Transactions on Networking}, vol.~28, no.~4, pp. 1562--1572, 2020.

\bibitem{wang2021optimal}
T.~Wang, X.~Wang, Z.~Wang, C.~Guo, W.~Moran, and M.~Zukerman, ``Optimal tree
  topology for a submarine cable network with constrained internodal latency,''
  \emph{Journal of Lightwave Technology}, vol.~39, no.~9, pp. 2673--2683, 2021.

\bibitem{eramo2018processing}
V.~Eramo and F.~Lavacca, ``Processing and bandwidth resource allocation in
  multi-provider {NFV} cloud infrastructures interconnected by elastic optical
  networks,'' in \emph{2018 20th International Conference on Transparent
  Optical Networks (ICTON)}.\hskip 1em plus 0.5em minus 0.4em\relax IEEE, 2018,
  pp. 1--6.

\bibitem{sethian1999level}
J.~A. Sethian, \emph{Level set methods and fast marching methods: evolving
  interfaces in computational geometry, fluid mechanics, computer vision, and
  materials science}.\hskip 1em plus 0.5em minus 0.4em\relax Cambridge
  university press, 1999, vol.~3.

\bibitem{gilbert1968steiner}
E.~N. Gilbert and H.~O. Pollak, ``{Steiner minimal trees},'' \emph{SIAM Journal
  on Applied Mathematics}, vol.~16, no.~1, pp. 1--29, 1968.

\bibitem{warme1998spanning}
D.~M. Warme, \emph{{Spanning trees in hypergraphs with applications to Steiner
  trees}}.\hskip 1em plus 0.5em minus 0.4em\relax University of Virginia
  Charlottesville, VA, 1998.

\bibitem{winter1997euclidean}
P.~Winter and M.~Zachariasen, ``{Euclidean Steiner minimum trees: An improved
  exact algorithm},'' \emph{Networks: An International Journal}, vol.~30,
  no.~3, pp. 149--166, 1997.

\bibitem{caleffi2014solution}
M.~Caleffi, I.~F. Akyildiz, and L.~Paura, ``On the solution of the steiner tree
  np-hard problem via physarum bionetwork,'' \emph{IEEE/ACM transactions on
  networking}, vol.~23, no.~4, pp. 1092--1106, 2014.

\bibitem{aharoni1998restricted}
E.~Aharoni and R.~Cohen, ``Restricted dynamic steiner trees for scalable
  multicast in datagram networks,'' \emph{IEEE/ACM transactions on Networking},
  vol.~6, no.~3, pp. 286--297, 1998.

\bibitem{GeoSteiner}
\BIBentryALTinterwordspacing
{David Warme, Pawel Winter, Martin Zachariasen}, ``{Software for Computing
  Steiner Trees},'' January 2017. [Online]. Available:
  \url{http://www.geosteiner.com/}
\BIBentrySTDinterwordspacing

\bibitem{smith1992find}
W.~D. Smith, ``{How to find Steiner minimal trees in Euclidean d-space},''
  \emph{Algorithmica}, vol.~7, no.~1, pp. 137--177, 1992.

\bibitem{fampa2008improved}
M.~Fampa and K.~M. Anstreicher, ``{An improved algorithm for computing Steiner
  minimal trees in Euclidean d-space},'' \emph{Discrete Optimization}, vol.~5,
  no.~2, pp. 530--540, 2008.

\bibitem{sun2020physarum}
Y.~Sun, D.~Rehfeldt, M.~Brazil, D.~Thomas, and S.~Halgamuge, ``A
  physarum-inspired algorithm for minimum-cost relay node placement in wireless
  sensor networks,'' \emph{IEEE/ACM Transactions on Networking}, vol.~28,
  no.~2, pp. 681--694, 2020.

\bibitem{wang2018application}
Z.~Wang, Q.~Wang, B.~Moran, and M.~Zukerman, ``Application of the fast marching
  method for path planning of long-haul optical fiber cables with shielding,''
  \emph{IEEE Access}, vol.~6, pp. 41\,367--41\,378, 2018.

\bibitem{wang2019terrain}
Z.~Wang, Q.~Wang, B.~Moran, and M.~Zukerman, ``Terrain constrained path planning for long-haul cables,''
  \emph{Optics express}, vol.~27, no.~6, pp. 8221--8235, 2019.

\bibitem{eriksson2013applied}
K.~Eriksson, D.~Estep, and C.~Johnson, \emph{Applied mathematics: Body and
  soul: Volume 1: Derivatives and geometry in IR3}.\hskip 1em plus 0.5em minus
  0.4em\relax Springer Science \& Business Media, 2013.

\bibitem{GMRT}
\BIBentryALTinterwordspacing
``{Global Multi-Resolution Topography Data Synthesis},'' October 2019.
  [Online]. Available: \url{https://www.gmrt.org/}
\BIBentrySTDinterwordspacing

\end{thebibliography}

\end{document}